\documentclass[reprint,amssymb, amsmath,twocolumn, aps, superscriptaddress, showpacs, footinbib, pra, longbibliography	]{revtex4-1}
\pdfoutput=1
\usepackage{graphicx} 
\usepackage{amsmath}
\usepackage{graphicx,epstopdf}

\newcommand{\be}{\begin{equation}}
\newcommand{\ee}{\end{equation}}
\newcommand{\bea}{\begin{eqnarray}}
\newcommand{\eea}{\end{eqnarray}}
\newcommand{\bse}{\begin{subequations}}
\newcommand{\ese}{\end{subequations}}
\newcommand{\kv}{\mathbf k}

\newcommand{\eps}{\varepsilon}
\begin{document}

\title{Commensurate and incommensurate magnetic order\\ in spin-1 chains stacked on the triangular lattice in Li$_{2}$NiW$_{2}$O$_{8}$}
\author{K. M. Ranjith}
\affiliation{School of Physics, Indian Institute of Science
Education and Research Thiruvananthapuram-695016, India}
\author{R.~Nath}
\email{rnath@iisertvm.ac.in}
\affiliation{School of Physics, Indian Institute of Science
Education and Research Thiruvananthapuram-695016, India}
\author{M.~Majumder}
\affiliation{Max Planck Institute for Chemical Physics of Solids, 01187 Dresden, Germany}
\author{D.~Kasinathan}
\affiliation{Max Planck Institute for Chemical Physics of Solids, 01187 Dresden, Germany}
\author{M.~Skoulatos}
\affiliation{Heinz Maier-Leibnitz Zentrum (MLZ) and Physics Department E21, Technische Universit\"at M\"unchen, 85748 Garching, Germany}
\affiliation{Laboratory for Neutron Scattering and Imaging, Paul Scherrer Insitut, CH-5232 Villigen PSI, Switzerland}
\author{L.~Keller}
\affiliation{Laboratory for Neutron Scattering and Imaging, Paul Scherrer Insitut, CH-5232 Villigen PSI, Switzerland}
\author{Y.~Skourski}
\affiliation{Dresden High Magnetic Field Laboratory, Helmholtz-Zentrum Dresden-Rossendorf, 01314 Dresden, Germany}
\author{M.~Baenitz}
\affiliation{Max Planck Institute for Chemical Physics of Solids, 01187 Dresden, Germany}
\author{A.~A.~Tsirlin}
\email{altsirlin@gmail.com}
\affiliation{National Institute of Chemical Physics and Biophysics, 12618 Tallinn, Estonia}
\affiliation{Experimental Physics VI, Center for Electronic Correlations and Magnetism, Institute of Physics, University of Augsburg, 86135 Augsburg, Germany}

\date{\today}
\begin{abstract}
We report thermodynamic properties, magnetic ground state, and microscopic magnetic model of the spin-1 frustrated antiferromaget Li$_{2}$NiW$_{2}$O$_{8}$ showing successive transitions at $T_{\rm N1}\simeq 18$\,K and $T_{\rm N2}\simeq 12.5$\,K in zero field. Nuclear magnetic resonance and neutron diffraction reveal collinear and commensurate magnetic order with the propagation vector $\kv=(\frac12,0,\frac12)$ below $T_{\rm N2}$. The ordered moment of 1.8\,$\mu_B$ at 1.5\,K is directed along $[0.89(9),-0.10(5),-0.49(6)]$ and matches the magnetic easy axis of spin-1 Ni$^{2+}$ ions, which is determined by the scissor-like distortion of the NiO$_6$ octahedra. Incommensurate magnetic order, presumably of spin-density-wave type, is observed in the region between $T_{\rm N2}$ and $T_{\rm N1}$. Density-functional band-structure calculations put forward a three-dimensional spin lattice with spin-1 chains running along the $[01\bar 1]$ direction and stacked on a spatially anisotropic triangular lattice in the $ab$ plane. We show that the collinear magnetic order in Li$_2$NiW$_2$O$_8$ is incompatible with the triangular lattice geometry and thus driven by a pronounced easy-axis single-ion anisotropy of Ni$^{2+}$. 
\end{abstract}
\maketitle

\section{Introduction}
Triangular-lattice antiferromagnets entail geometrical frustration of exchange couplings and show rich physics with a variety of ordered phases depending on temperature and applied magnetic field~\cite{collins1997,starykh2015}. Magnetic order on the regular triangular lattice with isotropic (Heisenberg) exchange couplings is non-collinear $120^{\circ}$ type even in the quantum case of spin-$\frac12$~\cite{jolicoeur1989,miyake1992,chubukov1994,capriotti1999,white2007}. In real materials, non-zero interplane couplings stabilize this order at finite temperatures up to the N\'eel temperature $T_{\rm N}$~\cite{kawamura1984}. However, exchange anisotropy, which is also inherent to real materials, will often render the magnetic behavior far more complex. 

In triangular antiferromagnets with easy-axis anisotropy, the 120$^{\circ}$ state is often preceded by a collinear state stabilized by thermal fluctuations~\cite{miyashita1985,melchy2009}. Zero-field measurements will then show two transitions with the collinear state formed below $T_{\rm N1}$ and the 120$^{\circ}$ state formed below $T_{\rm N2}<T_{\rm N1}$. This physics has been well documented experimentally in CsNiCl$_3$~\cite{clark1972,kadowaki1987}, RbMnI$_3$~\cite{ajiro1990,harrison1991}, and other triangular antiferromagnets. On the other hand, the intermediate-temperature collinear phase is unstable in systems with easy-plane anisotropy~\cite{miyashita1986}. Such systems will typically show only one magnetic transition in zero field, with the paramagnetic phase transforming directly into the $120^{\circ}$-ordered state below $T_{\rm N}$.

\begin{figure}
\includegraphics{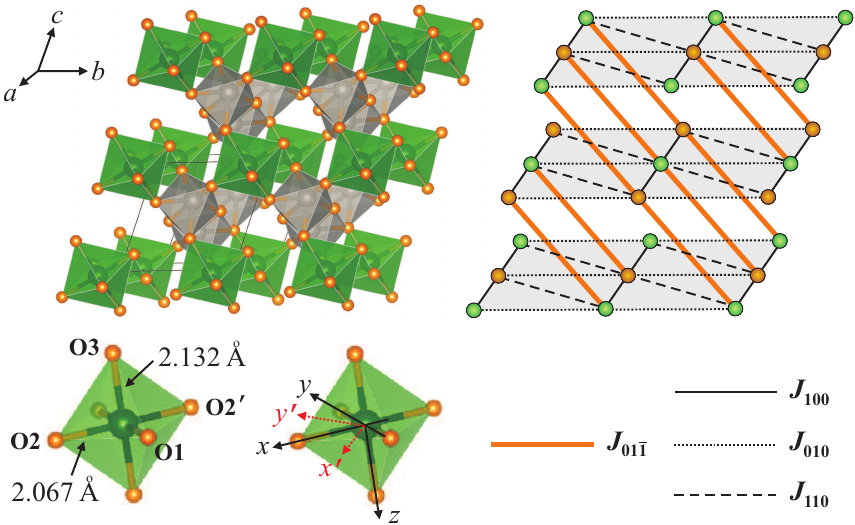}
\caption{\label{fig:structure}
(Color online) Crystal structure of Li$_2$NiW$_2$O$_8$ (left) and its spin lattice (right) featuring spin chains ($J_{01\bar 1}$) stacked on the triangular lattice in the $ab$ plane. Different colors denote opposite spin directions in the commensurate magnetic configuration below $T_{\rm N2}$ (see also Fig.~\ref{fig:refinement}). The Li atoms have been removed for clarity. The NiO$_6$ octahedra are shown in green, and their local coordinate frames $xyz$ and $x'y'z$ are also depicted (see Sec.~\ref{sec:dft} for details). \texttt{VESTA} software~\cite{vesta} was used for crystal structure visualization.
}
\end{figure}

Recently, Li$_2$NiW$_2$O$_8$ (Fig.~\ref{fig:structure}) was proposed as a triangular-like antiferromagnet showing two subsequent magnetic transitions at $T_{\rm N1}\simeq 18$\,K and $T_{\rm N2}\simeq 12.5$\,K in zero field~\cite{karna2015,panneer2015}. Below $T_{\rm N2}$, collinear and commensurate antiferromagnetic order was reported~\cite{karna2015}, in contrast to other triangular antiferromagnets~\cite{clark1972,kadowaki1987,ajiro1990,harrison1991}. The nature of the intermediate magnetic state between $T_{\rm N1}$ and $T_{\rm N2}$ and the origin of the collinear order below $T_{\rm N2}$ remain largely unclear. 

In the following, we report an independent study of Li$_2$NiW$_2$O$_8$ including thermodynamic properties, nuclear magnetic resonance, and neutron scattering, as well as a microscopic magnetic model obtained from density-functional (DFT) band-structure calculations. We revise the model of the collinear magnetic structure that is formed below $T_{\rm N2}$ and elucidate its origin by evaluating both isotropic exchange couplings and single-ion anisotropy. Our data reveal a consistent easy-axis anisotropy scenario that is verified experimentally and rationalized microscopically. We further demonstrate incommensurate nature of the magnetic order in the temperature range between $T_{\rm N2}$ and $T_{\rm N1}$, and discuss its possible origin. We also probe spin dynamics in both commensurate and incommensurate phases of Li$_2$NiW$_2$O$_8$.

\section{Methods}
Polycrystalline sample of Li$_{2}$NiW$_{2}$O$_{8}$ was prepared by the conventional solid-state
reaction technique using Li$_{2}$CO$_{3}$ (Aldich, $99.999$\%), NiO (Aldich, $99.999$\%), and WO$_{3}$ (Aldich, $99.999$\%) as starting materials. Stoichiometric mixtures of the reactants were ground and fired at 600$^{\circ}$C for 24 hours to allow decarbonation and then at 650$^{\circ}$C for 48 hours with one intermediate grinding. Phase purity of the sample was confirmed by powder x-ray diffraction (XRD) experiment (PANalytical, Empyrean) with Cu K$\alpha$ radiation ($\lambda_{\rm av}$ = 1.5406\,\r A). The data were refined using \texttt{FullProf}~\cite{RodriguezCarvajal55}.

Magnetization ($M$) and magnetic susceptibility ($\chi = M/H$) data were collected as a function of temperature ($T$) and magnetic field ($H$) using a SQUID magnetometer (Quantum Design, MPMS). Heat capacity ($C_{\rm p}$) measurement was performed as a function of $T$ and $H$ on a sintered pellet using the relaxation technique in the PPMS (Quantum Design). High-field magnetization data up to 60\,T were collected using a pulsed magnet installed at the Dresden High Magnetic Field Laboratory. Details of the experimental procedure can be found elsewhere~\cite{Tsirlin132407}. Our high field data were scaled with the magnetization data measured using SQUID magnetometer up to 7\,T.

Temperature dependent nuclear magnetic resonance (NMR) experiments were carried out using pulsed NMR technique on the $^7$Li nucleus which has a nuclear spin \mbox{$I=\frac32$} and gyromagnetic ratio $\bar{\gamma_{N}} = \gamma_{\rm N}/2\pi$ = 16.546\,MHz/T. Spectra were obtained either by Fourier transform of the NMR echo signal at two different fixed fields $H$ = 1.5109\,T and 0.906\,T or by sweeping the magnetic field at the corresponding fixed frequencies of 24.79\,MHz and 15\,MHz. $^7$Li spin-lattice relaxation rate $1/T_1$ was measured using a conventional saturation pulse sequence. Spin-spin relaxation rate 1/$T_2$ was carried out using a standard $\pi$/2-$\tau$-$\pi$ pulse sequence.

Neutron diffraction data were collected on the DMC diffractometer at SINQ (PSI, Villigen) operating at the wavelength of 2.45\,\r A. Data analysis was performed using the \texttt{JANA2006} software~\cite{jana2006}.

Exchange couplings in Li$_2$NiW$_2$O$_8$ were obtained from density-functional band-structure calculations performed in the \texttt{FPLO} code~\cite{fplo} using the experimental crystal structure and local-density approximation (LDA) exchange-correlation potential~\cite{pw92}. Correlation effects in the Ni $3d$ shell were taken into account on the mean-field level using the LSDA+$U$ approach with the fully-localized-limit double-counting correction scheme, on-site Coulomb repulsion $U_d=8$\,eV, and Hund's exchange $J_d=1$\,eV~\cite{anisimov1999}.

Magnetic susceptibility of the spin-1 chain was obtained from quantum Monte-Carlo simulations performed in the \texttt{looper} algorithm~\cite{loop} of the \texttt{ALPS} simulation package~\cite{alps}. A finite lattice with $L=32$ sites and periodic boundary conditions was used.

\section{Results}
\subsection{Crystallography}
Initial parameters for the structure refinement from the powder XRD data were taken from Ref.~\onlinecite{Vega3871}.
The obtained best fit parameters listed in Table~\ref{refinement} are in close agreement with the previous reports~\cite{Vega3871,karna2015}.
\begin{table}[ptb]
\caption{Crystal structure data for Li$_{2}$NiW$_{2}$O$_{8}$ at room temperature (Triclinic structure with space group $P\bar{1}$). Refined lattice parameters are $a = 4.9039(1)$\,\AA, $b = 5.5974(1)$\,\AA,
$c = 5.8364(1)$\,\AA, $\alpha = 70.888(1)^{\circ}$, $\beta = 88.5374(9)^{\circ}$, and $\gamma = 115.4419(8)^{\circ}$ which are comparable to the previously reported values~\cite{karna2015,Vega3871}.\footnote{Note that $c=5.4803$\,\r A reported in Ref.~\onlinecite{Vega3871} is due to misprint.} Our fit yields $\chi^2\simeq$ 7.6. Listed are the Wyckoff positions and the refined atomic coordinates ($x/a$, $y/b$, and $z/c$) for each atom.}
\label{refinement}
\begin{ruledtabular}
\begin{tabular}{ccccc}
Atom & Site & $x/a$ & $y/b$ & $z/c$ \\\hline
Li & $2i$ & $0.023(5) $ & $0.070(3)$ & $0.224(4)$ \\
Ni & $1d$ & $\frac12$ & $0$ & $0$ \\
O1 & $2i$ & $-0.029(3)$ & $0.287(2)$ & $0.931(2)$ \\
O2 & $2i$ & $0.512(3)$ & $0.718(2)$ & $0.785(2)$ \\
O3 & $2i$ & $0.555(3)$ & $0.699(2)$ & $0.356(2)$ \\
O4 & $2i$ & $0.045(3)$ & $0.692(3)$ & $0.572(2)$ \\
W  & $2i$ & $0.2604(3)$ & $0.531(3)$ & $0.6669(3)$ \\
\end{tabular}
\end{ruledtabular}
\end{table}

The crystal structure of Li$_2$NiW$_2$O$_8$ (Fig.~\ref{fig:structure}) features NiO$_6$ octahedra connected via WO$_6$ octahedra. The NiO$_6$ octahedra are weakly distorted with two longer Ni--O distances of 2.132\,\r A and four shorter distances of $2.04-2.07$\,\r A. The WO$_6$ octahedra are distorted as well and feature one short W--O distance of 1.77\,\r A, as typical for W$^{6+}$. Triclinic symmetry of the crystal structure results in a large number of nonequivalent superexchange pathways. Three shortest Ni--Ni distances are found in the $ab$ plane forming a triangular lattice, although no pronounced two-dimensionality can be expected on crystallographic grounds, because the shortest Ni--Ni distance along $c$ is only slightly longer than those in the $ab$ plane, 5.81\,\r A vs. $5.60-5.64$\,\r A (see also Table~\ref{tab:exchange}). In fact, the spin lattice of Li$_2$NiW$_2$O$_8$ turns out to be even more complex. While it preserves triangular features in the $ab$ plane, the leading exchange coupling is out of plane and oblique to the triangular layers (Fig.~\ref{fig:structure}, right). Details of magnetic interactions in Li$_2$NiW$_2$O$_8$ are further discussed in Sec.~\ref{sec:dft}.

\subsection{Magnetization}
\begin{figure}
\includegraphics{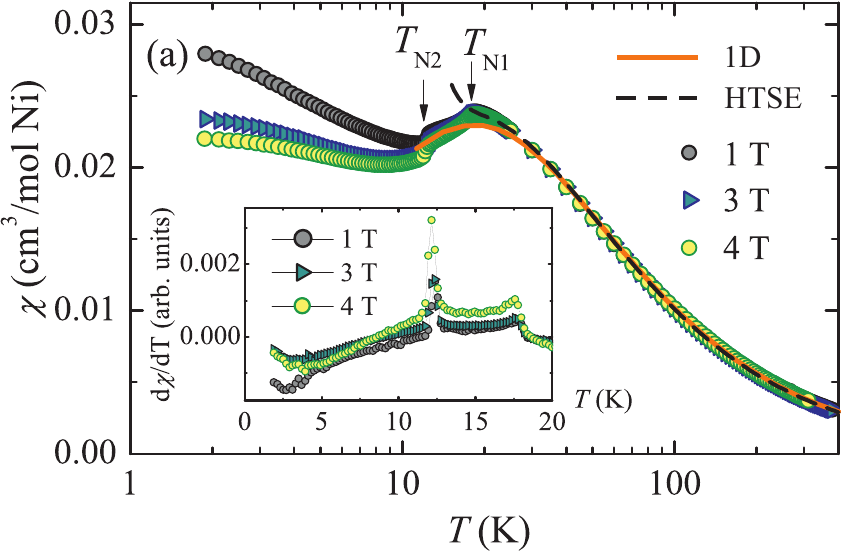}
\caption{\label{chi} Magnetic susceptibility of Li$_{2}$NiW$_{2}$O$_{8}$ as a function of temperature at different applied fields. The lines are fits as described in Sec.~\ref{sec:fits}. Inset: $d\chi/dT$ versus temperature measured at different fields in the low-$T$ regime highlighting the transition anomalies.}
\end{figure}

Magnetic susceptibility $\chi(T)$ data for Li$_{2}$NiW$_{2}$O$_{8}$ measured as a function of temperature at different applied fields are shown in Fig.~\ref{chi}. The susceptibility increases with decreasing temperature in a Curie-Weiss manner and shows a peak around $T_{\rm N1}\sim$ 18\,K and a change in slope at $T_{\rm N2}\sim$ 12.5\,K, suggesting that there are two magnetic transitions at low temperatures. These two transitions are more pronounced in the d$\chi$/d$T$ vs. $T$ plot [Inset of Fig.~\ref{chi}]. No broad maximum associated with the magnetic short-range order was observed above $T_{\rm N1}$. The field-dependence of the low temperature tail indicates an impurity contribution.

The $\chi(T)$ data at high temperatures were fitted by the sum of a temperature-independent contribution $\chi_0$ and the Curie-Weiss (CW) law:
\begin{equation}
 \chi=\chi_0+\frac{C}{T+\theta_{\rm CW}}.
\end{equation}
$\chi_0$ consists of diamagnetism of the core electron shells ($\chi_{\rm core}$) and Van-Vleck paramagnetism ($\chi_{\rm VV}$) of the open shells of the Ni$^{2+}$ ions present in the sample. In the CW law, the $\theta_{\rm CW}$ and $C$ stand for Curie-Weiss temperature and Curie constant, respectively.
Our fit to the 1\,T data (not shown) in the high temperature regime (175~K to 380~K) yields $\chi_0\simeq -8.738\times 10^{-6}$\,cm$^3$/mol, $C\simeq 1.24$\,cm$^3$\,K\,mol$^{-1}$, and $\theta_{\rm CW}\simeq$ 20\,K. The positive value of $\theta_{\rm CW}$ suggests that the dominant interactions are antiferomagnetic in nature. The value of $C$ yields an effective moment $\mu_{\rm eff}\simeq 3.15$\,$\mu_{\rm B}$/Ni which corresponds to a Land\'e $g$-factor $g\simeq$ 2.23, typical for the Ni$^{2+}$ ion~\cite{Hwang257205}. Adding core diamagnetic susceptibilities for individual ions ($\chi_{{\rm Li}^+}=-0.6\times 10^{-6}$\,cm$^3$/mol, $\chi_{{\rm Ni}^{2+}}=-12\times 10^{-6}$\,cm$^3$/mol, $\chi_{{\rm W}^{6+}}=-13\times 10^{-6}$\,cm$^3$/mol, and $\chi_{{\rm O}^{2-}}=-12\times 10^{-6}$\,cm$^3$/mol~\cite{Selwood1956}), the total $\chi_{\rm core}$ was calculated to be $-1.35\times 10^{-4}$\,cm$^3$/mol. The Van Vleck contribution for Li$_{2}$NiW$_{2}$O$_{8}$ was thus estimated by subtracting $\chi_{\rm core}$ from $\chi_{0}$ producing $\chi_{\rm VV}\sim 12.6\times 10^{-5}$\,cm$^3$/mol.

\begin{figure}
\includegraphics{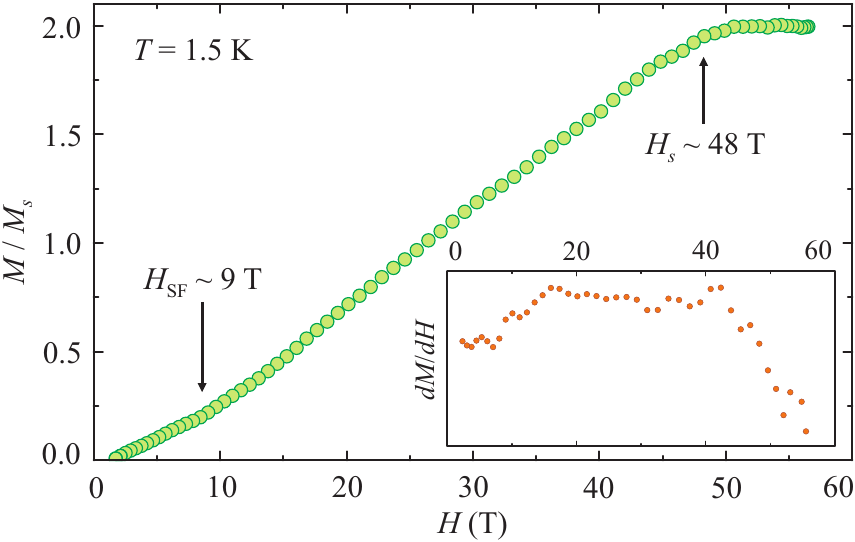}
\caption{\label{mh2} High-field magnetization curve measured at 1.5\,K in pulsed magnetic field. Inset: $dM/dH$ vs. $H$.}
\end{figure}
The pulsed-field magnetization data are shown in Fig.~\ref{mh2}. They exhibit two features: i) a change in the slope around $H_{\rm SF}\simeq 9$\,T, and ii) saturation around $H_s\simeq 48$\,T. The former feature is reminiscent of a spin-flop transition. Field derivative of the magnetization does not reveal any additional features in the region between 9\,T and 48\,T. We note, however, that narrow magnetization plateaus, which are not unexpected in triangular antiferromagnets~\cite{starykh2015}, may be smeared out by thermal fluctuations and anisotropy. Therefore, measurements at lower temperatures and on a single crystal are desirable to detect possible field-induced states in Li$_2$NiW$_2$O$_8$.

\subsection{Heat capacity}
\begin{figure}
\includegraphics{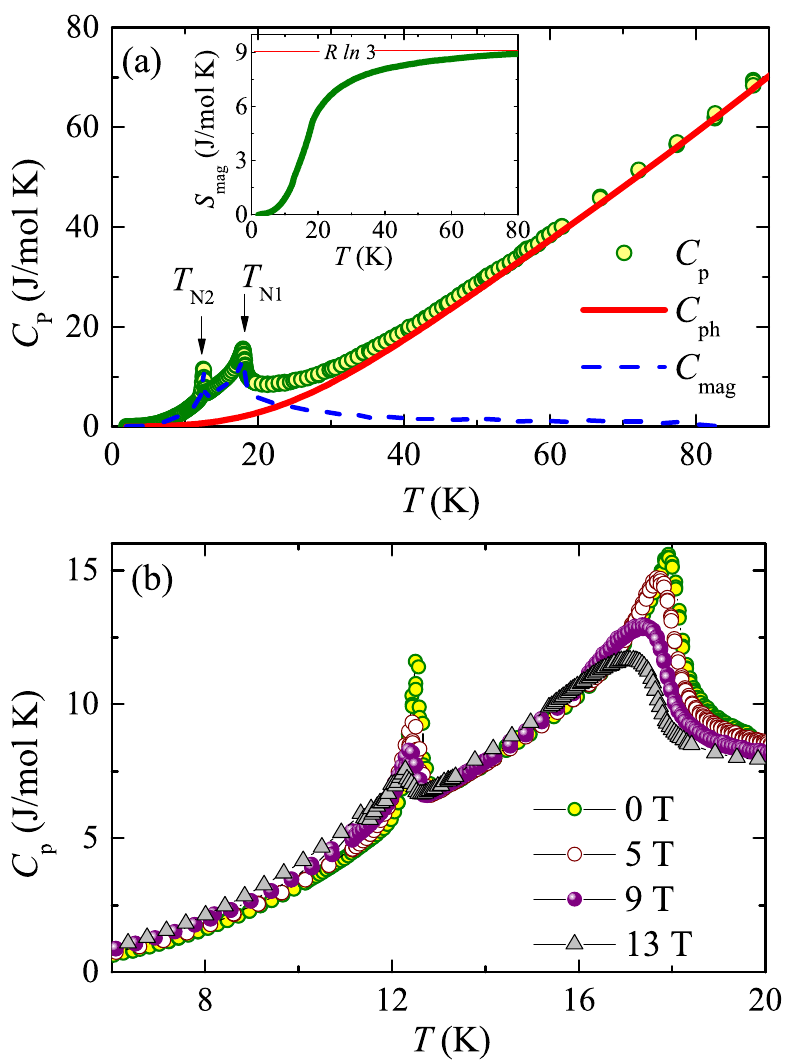}
\caption{\label{cp} (a) Temperature dependence of heat capacity measured at zero field for Li$_{2}$NiW$_{2}$O$_8$. The open circles are the raw data, the solid line is the phonon contribution $C_{\rm ph}$, and the dashed line represents the magnetic contribution $C_{\rm mag}$. Inset shows the magnetic entropy $S_{\rm mag}$ as a function of $T$. The horizontal line is the value $S_{\rm mag} = R\ln 3$ expected for Ni$^{2+}$ spins. (b) Heat capacity of Li$_{2}$NiW$_{2}$O$_8$ measured in different applied magnetic fields.}
\end{figure}
Heat capacity $C_{\rm p}$ as a function of temperature measured at zero field is presented in Fig.~\ref{cp}(a). At high temperatures, $C_{\rm p}$ is completely dominated by the contribution of phonon excitations. At low temperatures, it shows two $\lambda$-type anomalies at $T_{\rm N1}\sim$ 18~K and $T_{\rm N2}\sim$ 12.5~K indicative of two magnetic phase transitions, which are consistent with the $\chi(T)$ measurements. Below $T_{\rm N2}$, $C_{\rm p}(T)$ decreases gradually towards zero.

In order to estimate the phonon part of the heat capacity $C_{\rm ph}(T)$, the $C_{\rm p}(T)$ data at high temperature were fitted by a sum of Debye contributions following the procedure adopted in Ref.~\onlinecite{Nath064422}. Finally, the high-$T$ fit was extrapolated down to 2.1\,K and the magnetic part of the heat capacity $C_{\rm {mag}}(T)$ was estimated by subtracting $C_{\rm {ph}}(T)$ from $C_{\rm {p}}(T)$ [see Fig.~\ref{cp}(a)].


In order to check the reliability of the fitting procedure, we calculated the total magnetic entropy ($S_{\rm mag}$) by integrating $C_{\rm mag}(T)/T$ between 2.1\,K and high-temperatures as
\begin{equation}
\label{smag}
S_{\rm{mag}}(T) = \int_{2.1\,\text{K}}^{T}\frac{C_{\rm{mag}}(T')}{T'}\,dT'.
\end{equation}
The resulting magnetic entropy is $S_{\rm{mag}}\simeq$ 8.95\,J\,mol$^{-1}$\,K$^{-1}$ at 80\,K. This value matches closely with the expected theoretical value [$S_{\rm mag} = R\ln(2S+1) \simeq 9.1$\,J\,mol$^{-1}$\,K$^{-1}$] for a spin-1 system.

To gain more information about the magnetic ordering, we measured $C_{\rm p}(T)$ in different applied magnetic fields [see Fig.~\ref{cp}(b)]. With increasing $H$ from 0\,T to 13\,T, both $T_{\rm N1}$ and $T_{\rm N2}$ shifted towards lower temperatures suggesting an antiferromagnetic nature for both transitions. In addition to this, the intensities of the peaks are also decreasing with increasing $H$, which is likely due to the redistribution of magnetic entropy towards high temperatures.

\subsection{$^7$Li NMR}
\subsubsection{$^7$Li NMR spectra above $T_{\rm N}$}
$^7$Li is a quadrupole nucleus having nuclear spin $I=\frac32$ for which one would expect two satellite lines along with the central line from $^7$Li NMR. Our experimental $^7$Li NMR spectrum consists of a narrow spectral line without any satellites and the central peak position remains unchanged with temperature, similar to that observed for Li$_2$CuW$_2$O$_8$~\cite{ranjith2015}. Above $T_{\rm N1}$, the NMR line width tracks the magnetic susceptibility nicely. From the linear slope of the line width vs magnetic susceptibility plot with temperature as an implicit parameter, the dipolar coupling constant was calculated to be $A_{\rm dip}\simeq 6.7 \times$10$^{22}$\,cm$^{-3}$. This value of $A_{\rm dip}$ corresponds to the dipolar field generated by the nuclei at a distance $r\simeq 2.5$\,\r A apart, which is of the correct order of magnitude for the average distance between the Li nuclei and Ni$^{2+}$ ions in Li$_2$NiW$_2$O$_8$.

\subsubsection{$^7$Li spin-lattice relaxation rate $1/T_1$}
 \begin{figure}
\includegraphics {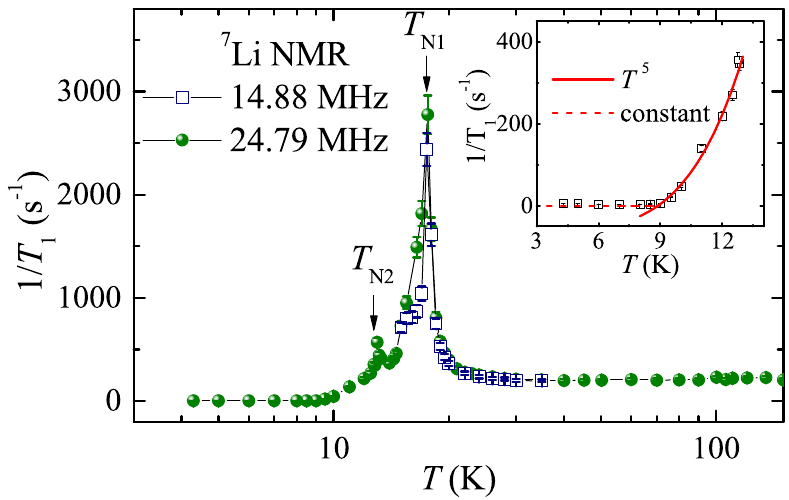}
\caption{\label{T1} Temperature-dependent $^7$Li spin-lattice relaxation rate, $1/T_1$, for Li$_2$NiW$_2$O$_8$ measured at two different frequencies of 14.88\,MHz and 24.79\,MHz. The downward arrows point to $T_{\rm N1}$ and $T_{\rm N2}$. Inset: 1/$T_1$ below $T_{\rm N2}\simeq$ 12.5\,K. The solid line represents the $1/T_1\propto T^5$ behavior and the dashed line corresponds to the $T$-independent behavior.}
\end{figure}
The decay of the longitudinal magnetization after a saturation pulse was fitted well by a single exponential function:
\begin{equation}
1-\frac{M(t)}{M_0} = Ae^{-t/T_1},
\label{expo1}
\end{equation}
where $M(t)$ is the nuclear magnetization at a time $t$ after the saturation pulse, and $M_0$ is the equilibrium value of the magnetization. At low temperatures ($T\leq 10$\,K), the recovery curve deviates from single exponential behavior but fits well to a double exponential (consisting of a short and a long component) function:
\begin{equation}
1-\frac{M(t)}{M_0} = A_1e^{-t/T_{1L}}+A_2e^{-t/T_{1S}},
\label{expo2}
\end{equation}
where $T_{1L}$ and $T_{1S}$ represent the long and the short components of $T_{1}$, and $A_{1}$ and $A_2$ stand for their respective weight factors.
The long component 1/$T_{1L}$ of the double exponential fit matches with the 1/$T_1$ of the single exponential fit around 10\,K and are plotted as a function of temperature in Fig.~\ref{T1}. At high temperatures ($T\geq 40\,K$), 1/$T_1$ is almost temperature-independent, as expected in the paramagnetic regime~\cite{Moriya516}. With decrease in temperature, 1/$T_1$ increases slowly and shows a sharp peak at $T_{\rm N1}\simeq$ 17.6\,K and another weak anomaly at $T_{\rm N2}\simeq$ 12.5\,K associated with the two magnetic phase transitions. These results are consistent with our thermodynamic measurements. Below $T_{\rm N2}$, 1/$T_1$ decreases smoothly and becomes temperature-independent with a value nearly close to zero, below 9\,K. This type of temperature-independent 1/$T_1$ behavior in the ordered state is rather unusual but has been observed in the decorated Shastry-Sutherland lattice compound CdCu$_2$(BO$_3$)$_2$~\cite{Lee214416}. The origin of such a behaviour is not yet clear and requires further investigations.

\begin{figure}
\includegraphics {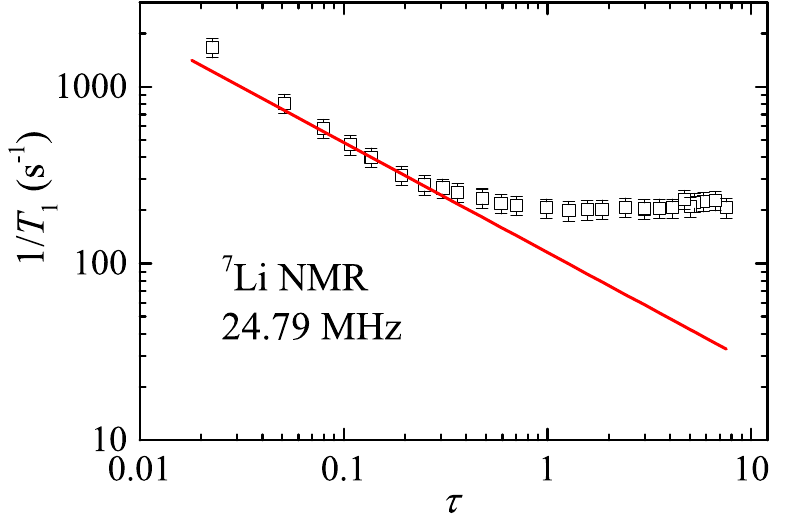}
\caption{\label{critical} $^7$Li 1/$T_1$ for Li$_{2}$NiW$_{2}$O$_{8}$ vs. the reduced temperature ($\tau$). The solid line is the fit by the power law, $1/T_1 \propto \tau^{-\gamma}$ with $\gamma\simeq$ 0.75 and a fixed $T_{\rm N1}\simeq$ 18\,K.}
\end{figure}
In order to analyze the critical behavior close to $T_{\rm N1}$, we have plotted 1/$T_1$ versus the reduced temperature $\tau$ = ($T-T_{\rm N1})/T_{\rm N1}$ above $T_{\rm N1}$ in Fig.~\ref{critical}. Usually in the critical regime (i.e., in the vicinity of $T_{\rm N}$) one would expect a critical divergence of 1/$T_1$ following the power law $1/T_1 \propto \tau^{-\gamma}$, where $\gamma$ is the critical exponent. When approaching $T_{\rm N}$ from above, the correlation length is expected to diverge and the fast electron spin fluctuations are slowed down. The value of $\gamma$ represents the universality class of the system that depends on the dimensionality and symmetry of the spin lattice, and on the type of magnetic interactions. According to the theoretical predictions, the $\gamma$ value is close to 0.3, 0.6, 0.8, and 1.5 for 3D Heisenberg, 3D Ising, 2D Heisenberg, and 2D Ising antiferromagnets, respectively~\cite{Benner1990}. Our 1/$T_1$ is fitted well by a power law for $\tau\leq 0.3$ giving $\gamma\simeq 0.60-0.75$. This would be consistent with either 3D Ising or 2D Heisenberg cases. The former option looks plausible, given the easy-axis anisotropy present in Li$_2$NiW$_2$O$_8$ and the 3D nature of the magnetic order formed below $T_{\rm N1}$.

In the magnetically ordered state ($T<T_{\rm N}$), the strong temperature dependence of 1/$T_1$ is a clear signature of the relaxation due to scattering of magnons by the nuclear spins~\cite{Belesi184408}. For $T\gg\Delta/k_{\rm B}$, 1/$T_1$ follows either a $T^3$ behavior or a $T^5$ behavior due to a two-magnon Raman process or a three-magnon process, respectively, where $\Delta/k_B$ is the energy gap in the spin-wave spectrum~\cite{Beeman359,Nath024431}. On the other hand, for $T\ll\Delta/k_{\rm B}$, it follows an activated behavior 1/$T_1 \propto T^2e^{-\Delta/k_{\rm B}T}$. As shown in the inset of Fig.~\ref{T1}(a), 1/$T_1$ for Li$_{2}$NiW$_{2}$O$_{8}$ below $T_{\rm N2}$ follows a $T^5$ behavior rather than a $T^3$ behavior, which ascertains that the relaxation is mainly governed by three-magnon process similar to that reported for spin-$\frac12$ square-lattice compound Zn$_2$VO(PO$_4$)$_2$~\cite{Yogi024413}, three-dimensional frustrated antiferromagnet Li$_{2}$CuW$_{2}$O$_{8}$~\cite{ranjith2015}, and decorated Shastry-Sutherland compound CdCu$_2$(BO$_3$)$_2$~\cite{Lee214416}.

We also measured 1/$T_1$ at two different temperatures, 15\,K and 22\,K, for different applied fields (not shown) and it was found to be field-independent, similar to our earlier observations for Li$_2$CuW$_2$O$_8$~\cite{ranjith2015}.

Spin-spin relaxation rate, 1/$T_2$ as a function of temperature (not shown) also shows two sharp peaks at $T_{\rm N1}\simeq 17.6$\,K and $T_{\rm N2}\simeq 12.5$\,K corresponding to the magnetic phase transitions as those observed in the 1/$T_1$ data.

\subsubsection{$^7$Li NMR spectra below $T_N$}
\begin{figure}
\includegraphics {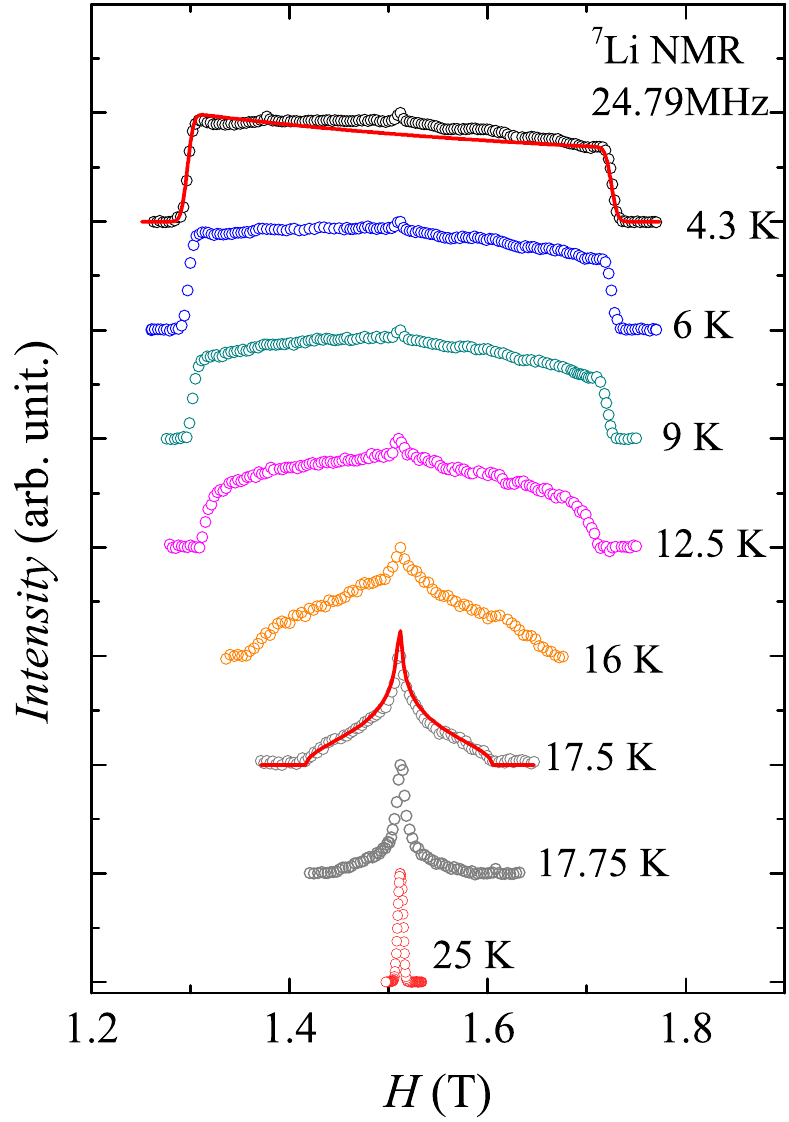}
\caption{\label{Sweep} Field-sweep $^7$Li NMR spectra measured at 24.79\,MHz in the low-temperature regime. Solid lines are the fits by Eq.~\eqref{convolv} at 4.3\,K, in the commensurately ordered state with $H_{\rm n} \simeq 0.213$\,T, and by Eq.~\eqref{incom} at 17.5\,K, in the incommensurately ordered state with $(H_{\rm n})_{\rm max} \simeq 0.094$\,T.}
\end{figure}
Figure~\ref{Sweep} shows field-sweep $^7$Li NMR spectra measured at 24.79\,MHz in the low-temperature regime. The NMR line broadens systematically with decreasing temperature. Below $T_{\rm N1}$, $^7$Li NMR spectra were found to broaden abruptly due to internal field in the ordered state and take a nearly triangular shape. The actual spectrum looks like a superposition of a broad background and a narrow central line at the zero-shift position. Upon lowering the temperature, this line broadening increases and the intensity of the central line decreases. Below the second transition $T_{\rm N2}\simeq$ 12.5\,K, the line broadening is drastic, and the line attains a nearly rectangular shape with the center of gravity at 1.512\,T, which is the resonance field for the $^7$Li nuclei. In order to reconfirm the line shape, we remeasured the spectra at a lower frequency of 14.88\,MHz. The width and shape of the NMR spectra were found to be identical with the spectra measured at 24.79\,MHz.

\begin{figure}
\includegraphics {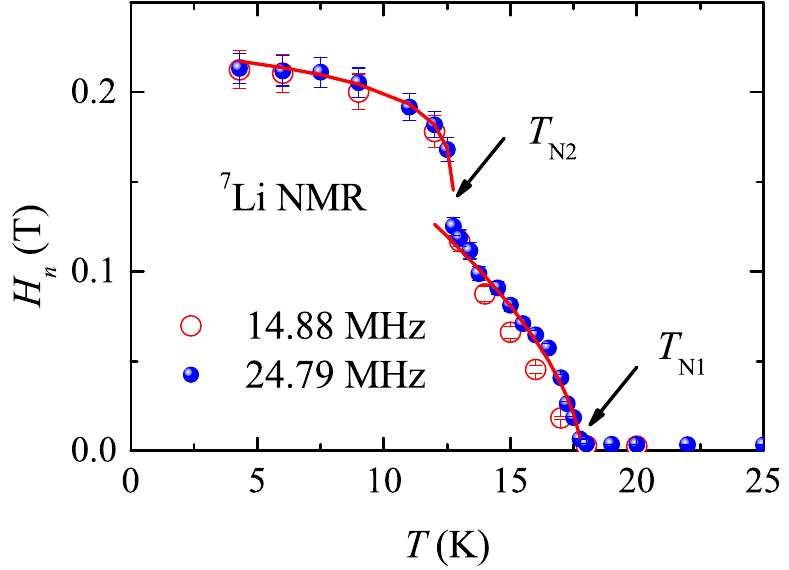}
\caption{\label{Hn} Temperature dependence of the internal field $H_n$ (\mbox{$T\leq T_{\rm N}$}) obtained from $^7$Li NMR spectra at two different frequencies. The solid lines guide the eye, and the arrows point to the transition points.}
\end{figure}
Our data reveal collinear and commensurate magnetic order in Li$_2$NiW$_2$O$_8$ already below $T_{\rm N2}$. In powder samples, the direction of the internal field is randomly distributed with respect to that of the applied field $H$. Therefore, one can express the NMR spectrum $f(H)$ as~\cite{Yamada1751,Kikuchi2660}:
\begin{equation}
f(H) \propto \frac{H^2 - H_{\rm n}^2 + H_{0}^2}{H_{\rm n}H^2},
\end{equation}
where $H_0= \omega/\bar{\gamma}_{N} =|H + H_n|$ is the central field, $\omega$ is the NMR frequency, $\bar{\gamma}_{\rm N}$ is the nuclear gyromagnetic ratio, and $H_{\rm n}$ is the internal field.
Two cutoff fields, $H_0+H_n$ and $H_0-H_n$, produce two sharp edges of the spectrum. In real materials, these sharp edges are usually smeared by the inhomogeneous distribution of the internal field. In order to take this effect into account, we convoluted $f(H)$ with a Gaussian distribution function $g(H)$:
\begin{equation}
 F(H) = \int f(H - H')g(H') dH',
 \label{convolv}
\end{equation}
 where the distribution function is taken as~\cite{Kikuchi2660}:
\begin{equation}
  g(H) = \frac{1}{\sqrt{2\pi\Delta H_{n}^2}} \exp\left(-\frac{(H - H_{n})^2}{2 \Delta H_n}\right).
\end{equation}
The spectra below $T_{\rm N2}$ are well fitted by Eq.~\eqref{convolv}, as shown in Fig.~\ref{Sweep}. This confirms the commensurate nature of the ordered state below $T_{\rm N2}\simeq 12.5$\,K.

Above $T_{\rm N2}$, the magnetic order changes drastically. The NMR spectra have a nearly triangular shape in this temperature range. Their maximum is at $\bar{\gamma}_{\rm N}H_0$, as shown in Fig.~\ref{Sweep}, similar to that reported for Cu$_3$B$_2$O$_6$~\cite{Sakurai024428}, (DIETSe)$_2$GaCl$_4$~\cite{Michioka042124}, and Cr$_{1-x}$T$_x$B$_2$ ($T$ = Mo and V)~\cite{Yoshimura709}. This triangular-like shape of the NMR spectra is a typical powder pattern of an incommensurate spin-density-wave (SDW) state~\cite{Kontani672}.

In an incommensurately ordered state, the powder NMR spectral shape function $f(x)$ can be written as~\cite{Kontani672}:
\begin{equation}
f(x) \propto \ln\left|\frac{1+\sqrt{1-x^2}}{x}\right|,
\label{incom}
\end{equation}
where $x=(H-\omega/\bar{\gamma}_{\rm N})/(H_{\rm n})_{\rm max}$, $-1\leq x\leq 1$, ($H_{\rm n})_{\rm max}$ is the maximum amplitude of the internal field, and sinusoidal modulation is assumed. As shown in Fig.~\ref{Sweep}, the experimental spectrum at $T = 17.5$\,K is well fitted by Eq.~\eqref{incom}, thus corroborating incommensurate magnetic order between $T_{\rm N1}$ and $T_{\rm N2}$ inferred from neutron diffraction (see below).

Temperature dependence of the internal field $H_{\rm n}$ extracted from the NMR spectra is presented in Fig.~\ref{Hn}. The internal field increases rapidly below about 18\,K and then develops the tendency of saturation towards low temperatures. Below 12.5\,K, $H_{\rm n}$ shows another sharp increase and then saturates below 5\,K, thus reflecting the two magnetic transitions.

We should notice that the central line (Fig.~\ref{Sweep}) does not disappear completely even at 4.3\,K. Persistence of this tiny central peak at 4.3~K could be due to small amount of defects and dislocations which is of course unavoidable in powder samples.


\subsection{Neutron diffraction}
\label{sec:neutron}
\begin{figure}
\includegraphics{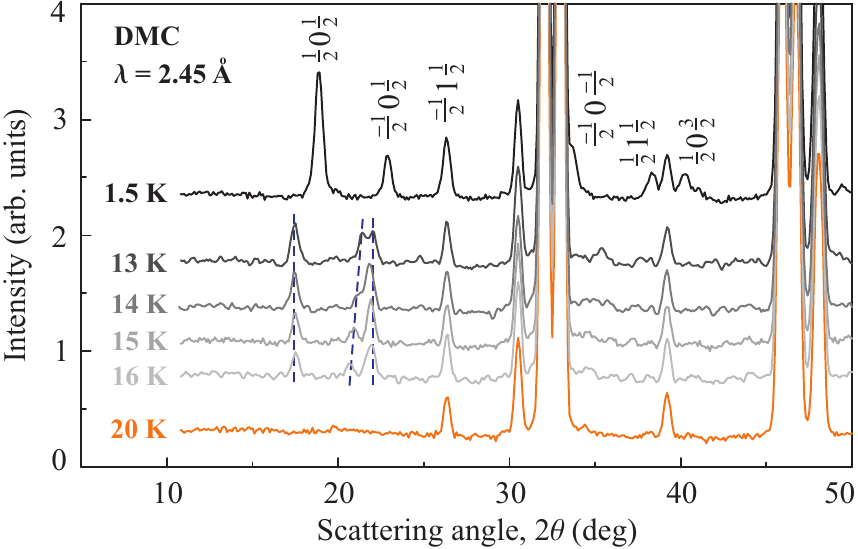}
\caption{\label{fig:neutron}
(Color online) Neutron diffraction data collected from Li$_2$NiW$_2$O$_8$ at different temperatures. The lower pattern shows nuclear scattering at 20\,K. The upper pattern (1.5\,K) corresponds to the commensurate phase below $T_{\rm N2}$. The indexes of the magnetic refections are also shown. The sequence of patterns collected between 13 and 16\,K illustrates temperature evolution of the incommensurate phase ($T_{\rm N2}<T<T_{\rm N1}$), with dashed lines tracking positions of individual magnetic reflections. Note that the patterns are offset for clarity. 
}
\end{figure}

Representative neutron powder diffraction patterns of Li$_2$NiW$_2$O$_8$ are shown in Fig.~\ref{fig:neutron}. Above $T_{\rm N1}$, only nuclear scattering is observed. It is consistent with the triclinic crystal structure reported in the literature~\cite{Vega3871}. Upon cooling below $T_{\rm N1}$, several weak magnetic reflections appear. Below $T_{\rm N2}$, these reflections abruptly disappear, and new magnetic reflections are observed at different angular positions. Data collected with a temperature step of 0.5\,K in the vicinity of $T_{\rm N2}$ (not shown) do not reveal coexistence of the two magnetically ordered phases, thus confirming a second-order nature for this transition.

\begin{figure}
\includegraphics{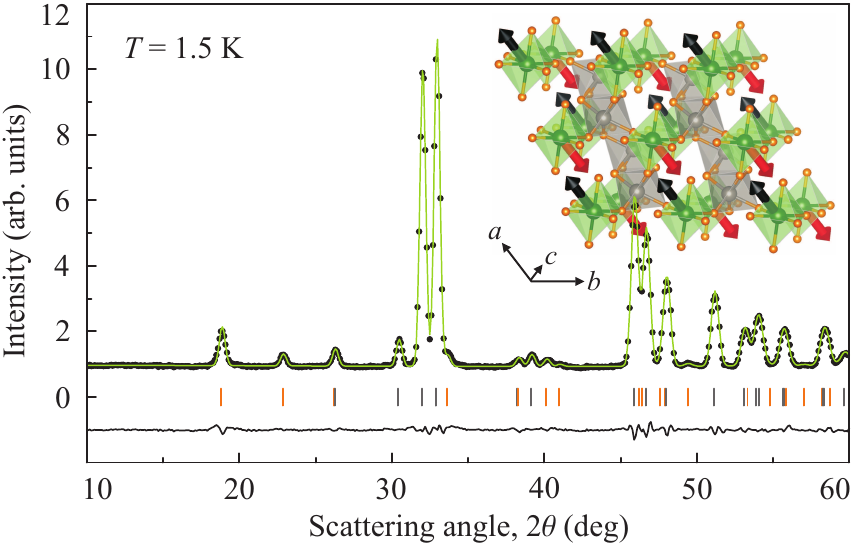}
\caption{\label{fig:refinement}
(Color online) Refinement of the magnetic and nuclear structures at 1.5\,K. The resulting spin arrangement is shown in the inset. Orange and gray tick marks denote the positions of the magnetic and nuclear peaks, respectively.
}
\end{figure}

Below $T_{\rm N2}$, all magnetic reflections can be indexed with the commensurate propagation vector $\kv=(\frac12,0,\frac12)$. Therefore, we find an AFM structure with antiparallel spins along $a$ and $c$ and parallel spins along $b$ (Fig.~\ref{fig:refinement}). At 1.5\,K, individual components of the magnetic moments are: $\mu_a=1.50(13)$\,$\mu_B$, $\mu_b=-0.18(9)$\,$\mu_B$, and $\mu_c=-0.85(11)$\,$\mu_B$. The ensuing ordered moment $\mu\simeq 1.8(1)$\,$\mu_B$ is slightly below 2.0\,$\mu_B$ expected for a \mbox{spin-1} ion and follows the direction of the Ni$^{2+}$ easy axis (see Sec.~\ref{sec:dft}). Upon heating from 1.5\,K to $T_{\rm N2}$, the ordered moment remains nearly constant [$\mu=1.7(1)$\,$\mu_B$ at 12\,K], in agreement with temperature dependence of the internal field probed by NMR (Fig.~\ref{Hn}).

A similar magnetic structure has been reported by Karna~\textit{et al.}~\cite{karna2015}, although with a very different direction of the magnetic moment. They found $\mu_a\simeq\mu_c$ and $\mu_b=0$, while in our case $\mu_a$ and $\mu_c$ have opposite signs. Our refinement is based on five well-resolved magnetic reflections, yet only two individual reflections could be resolved in Ref.~\onlinecite{karna2015}. Moreover, the spin direction found in our work is compatible with the magnetic easy axis of Ni$^{2+}$ in Li$_2$NiW$_2$O$_8$, while the one from Ref.~\onlinecite{karna2015} is not (see Sec.~\ref{sec:dft}).

In the temperature range between $T_{\rm N1}$ and $T_{\rm N2}$, only three magnetic reflections are observed (Fig.~\ref{fig:neutron}). Two of these reflections are temperature-independent, whereas the third one shifts to lower angles upon heating. This complex evolution of the magnetic reflections indicates that more than one propagation vector may be required to describe the intermediate-temperature phase in Li$_2$NiW$_2$O$_8$, and, unfortunately, no reliable solution of the magnetic structure could be found based on powder data. Importantly, though, none of the three reflections can be indexed with a commensurate propagation vector, thus indicating the incommensurate nature of the magnetic order, in agreement with the triangular shape of the NMR spectra in the temperature range between $T_{\rm N2}$ and $T_{\rm N1}$ (see Fig.~\ref{Sweep} and Sec.~\ref{sec:ordered}). 

\subsection{Microscopic magnetic model}
\label{sec:dft}
\subsubsection{Isotropic exchange couplings}
The relatively simple crystal structures of Li$_2$$M$W$_2$O$_8$ ($M$ = Cu, Ni, and Co)~\cite{Vega3871} host remarkably complex magnetic interactions. Recently, we reported a very intricate and strongly frustrated three-dimensional spin lattice of Li$_2$CuW$_2$O$_8$~\cite{ranjith2015a}, where magnetic interactions up to the Cu--Cu distance of 9.3\,\r A play crucial role. The spin lattice of the Ni compound is less involved, but also unusual. Individual magnetic interactions are obtained from LSDA+$U$ calculations~\footnote{LSDA+$U$ calculations converge to an insulating solution with the band gap of $3.2-3.4$\,eV.} as energy differences between collinear spin states~\cite{xiang2013}. The resulting exchange couplings $J_i$ parametrize the spin Hamiltonian:
\begin{equation}
 \hat H=\sum_{\langle ij\rangle}J_{ij}\mathbf S_i\mathbf S_j-\sum_i AS_{i\varkappa}^2,
\label{eq:ham}\end{equation}
where the summation is over lattice bonds $\langle ij\rangle$, and the local spin is $S=1$. The second term denotes single-ion anisotropy $A$ that will be discussed later in this section. Exchange couplings are labeled using crystallographic directions, where we take advantage of the fact that there is only one Ni atom per unit cell. 

\begin{figure}
\includegraphics{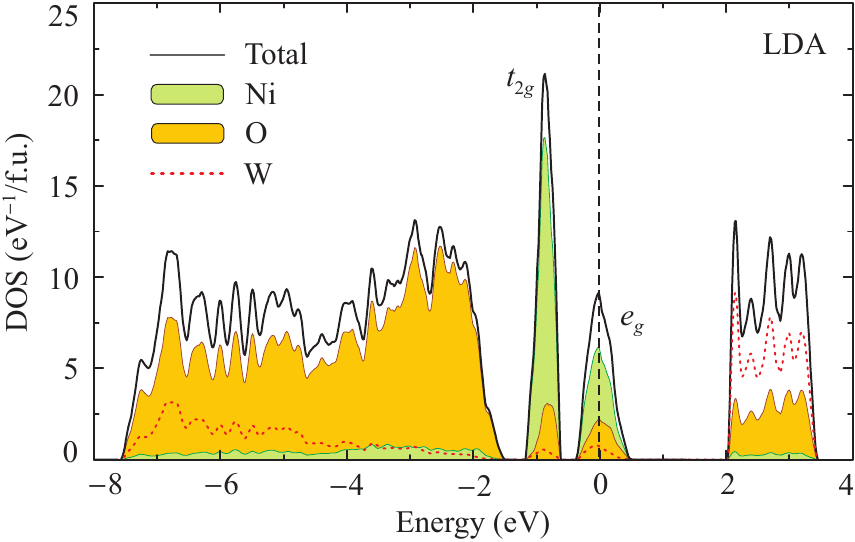}
\caption{\label{fig:dos}
(Color online) LDA density of states (DOS) for Li$_2$NiW$_2$O$_8$ featuring Ni $3d$ states around the Fermi level (zero energy), O $2p$ states below $-1.5$\,eV and W $5d$ states above 2\,eV. Note the splitting of the Ni $3d$ bands into $t_{2g}$ and $e_g$ complexes in the octahedral crystal field.
}
\end{figure}

The $J_i$'s listed in the last column of Table~\ref{tab:exchange} are obtained from LSDA+$U$ calculations. We find that all leading interactions in Li$_2$NiW$_2$O$_8$ are AFM. Additional microscopic insight into the origin of these couplings can be obtained from LDA band structure, where correlation effects are largely neglected. This metallic band structure (Fig.~\ref{fig:dos}) features well-defined Ni $3d$ bands that are split into $t_{2g}$ and $e_g$ complexes by the octahedral crystal field. The Fermi level bisects the half-filled $e_g$ states. Their tight-binding description yields hopping parameters $t_i^{\alpha\beta}$, where $\alpha\beta=x^2-y^2,3z^2-r^2$ are magnetic orbitals of Ni$^{2+}$. These hopping parameters listed in Table~\ref{tab:exchange} underlie the AFM superexchange $J_i$ in Li$_2$NiW$_2$O$_8$. 

The analysis of hopping parameters provides direct comparison to Li$_2$CuW$_2$O$_8$~\cite{ranjith2015a} with its spin-$\frac12$ Cu$^{2+}$ ions featuring $x^2-y^2$ as the only magnetic orbital. In both Cu and Ni compounds, same couplings $J_{100}$, $J_{010}$, and $J_{110}$ are active in the $ab$ plane, whereas $J_{\bar 110}$ corresponding to the longer Ni--Ni distance is negligible~\footnote{Note that $J_{110}$ and $J_{\bar 110}$ are swapped in the Cu and Ni compounds according to the different lattice angles $\gamma<90^{\circ}$ and $\gamma>90^{\circ}$, respectively~\cite{Vega3871}.}. On the other hand, the couplings between the $ab$ planes are remarkably different. Li$_2$CuW$_2$O$_8$ reveals the leading interplane coupling $J_{011}$ along with the somewhat weaker $J_{11\bar 1}$ and $J_{\bar 101}$. In contrast, Li$_2$NiW$_2$O$_8$ features $J_{0\bar 11}$ that is triggered by the hopping between the $3z^2-r^2$ orbitals and thus endemic to the Ni compound. The direct coupling along the $c$ axis ($J_{001}$) as well as other interplane couplings beyond $J_{0\bar 11}$ are much weaker than the three leading couplings in the $ab$ plane. Therefore, Li$_2$NiW$_2$O$_8$ lacks extraordinarily long-range couplings that are integral to the spin lattice of Li$_2$CuW$_2$O$_8$~\cite{ranjith2015a}.

\begin{table}
\caption{\label{tab:exchange}
Exchange couplings in Li$_2$NiW$_2$O$_8$. The Ni--Ni distances $d_{\text{Ni--Ni}}$ are in\,\r A, the hopping parameters $t_i^{\alpha\beta}$ are in\,meV, and exchange integrals $J_i$ are in\,K. The orbital indices are $\alpha=3z^2-r^2$ and $\beta=x^2-y^2$.
}
\begin{ruledtabular}
\begin{tabular}{c@{\hspace{2em}}c@{\hspace{2em}}rrr@{\hspace{2em}}r}
  & $d_{\text{Ni--Ni}}$ & 
\begin{tabular}{c} \\ $\alpha\alpha$ \end{tabular} &
\begin{tabular}{c} $t_i$ \\ $\alpha\beta$ \end{tabular} &
\begin{tabular}{c} \\ $\beta\beta$ \end{tabular} &
  $J_i$ \\ \hline
  $J_{100}$     & 4.907 & $-30$ & $-38$ & $-14$ & 3.1  \\
	$J_{010}$     & 5.602 & $-40$ &   32  &    9  & 2.1  \\
	$J_{110}$     & 5.642 &   22  &    7  & $-58$ & 2.6  \\
	$J_{001}$     & 5.810 &  $-1$ &  $-6$ &   17  & $-0.2$ \\
	$J_{01\bar 1}$ & 6.623 & $-111$ & $-6$ &   0  &  9.9 \\
\end{tabular}
\end{ruledtabular}
\end{table}

The spin lattice of Li$_2$NiW$_2$O$_8$ is quasi-1D. It entails the leading coupling $J_{01\bar 1}$ forming spin chains, and three weaker couplings forming triangular loops in the $ab$ plane (Fig.~\ref{fig:structure}). Therefore, this compound can be viewed as a system of spin-1 chains stacked on the triangular lattice, similar to, e.g., CsNiCl$_3$~\cite{clark1972,kadowaki1987}. The coupling geometry is in line with previous results, and the absolute values of $J_i$ are rather similar to those reported in Ref.~\onlinecite{panneer2015}, while exchange couplings in Ref.~\onlinecite{karna2015} are roughly 4 times larger. A closely related exchange topology has been reported for Li$_2$CoW$_2$O$_8$ as well~\cite{panneer2014}. On the other hand, Li$_2$CuW$_2$O$_8$ is very different~\cite{ranjith2015a}. It shows a 1D feature in the $ab$ plane along with an intricate network of frustrated exchange couplings in all three dimensions.

\subsubsection{Comparison to the experiment}
\label{sec:fits}
Our computed exchange couplings compare well to the experimental data. The Curie-Weiss temperature is a sum of exchange couplings at the magnetic site: 
\begin{equation}
 \theta=\frac{S(S+1)}{3}\sum_i z_iJ_i,
\end{equation}
where $z_i=2$ is the number of couplings per site. We obtain $\theta_{\rm CW}=23.6$\,K that compares well to the experimental $\theta_{\rm CW}\simeq 20$\,K. 

The saturation field reflects the energy difference between the ground-state spin configuration and the ferromagnetic state. Using the classical energy of the $\kv=(\frac12,0,\frac12)$ AFM state, we find:
\begin{equation}
  H_s=2S(2J_{100}+2J_{110}+2J_{01\bar 1}) k_B/(g\mu_B)=46.6\,\text{T},
\end{equation}
which is again in good agreement with the experimental $H_s\simeq 48$\,T. We further note that the hierarchy of exchange couplings in Table~\ref{tab:exchange} is compatible with the $\kv=(\frac12,0,\frac12)$ order observed below $T_{\rm N2}$. The AFM coupling $J_{100}$ imposes AFM order along the $a$ direction. On the other hand, the AFM diagonal coupling $J_{110}$ is stronger than $J_{010}$. Therefore, the magnetic order is AFM along $[110]$ and thus FM along $[010]$. Likewise, the AFM coupling $J_{0\bar 11}$ combined with the FM order along $b$ results in the AFM order along $c$, and the $\kv=(\frac12,0,\frac12)$ state ensues. 

The quantitative description of thermodynamic properties can be extended by fitting temperature dependence of the magnetic susceptibility (Fig.~\ref{chi}). To this end, we first consider the model of an isolated spin-1 chain that describes the experimental data down to 30\,K resulting in $J_{\rm 1D}\simeq 14.1$\,K and $g=2.23$. Unfortunately, the susceptibility of the full 3D model can not be obtained from quantum Monte Carlo simulations because of the sign problem. Therefore, we resort to the high-temperature series expansion (HTSE)~\cite{htse,*htsecode}. To reduce the number of variable parameters, we fixed the ratios of $J_{100}$, $J_{010}$, and $J_{110}$. The fit describes the data down to 20\,K, i.e., nearly down to $T_{\rm N1}$, arriving at $J_{01\bar 1}\simeq 13.0$\,K, $J_{100}\simeq 0.5$\,K, and $g=2.21$ in reasonable agreement with the DFT results in Table~\ref{tab:exchange}. Note, however, that the estimate of $J_{100}$ (and thus of all couplings in the $ab$ plane) is rather sensitive to the temperature range of the fit and, additionally, the effect of the single-ion anisotropy can not be included in the HTSE available so far~\cite{htse}.

\begin{table}
\caption{\label{tab:orbitals}
Orbital energies $\eps_i$ in Li$_2$NiW$_2$O$_8$ calculated for two coordinate frames shown in Fig.~\ref{fig:structure}. The energies are in\,eV. The contributions to the single-ion anisotropy $A_x$, $A_y$, and $A_z$ (in\,K) are calculated using Eqs.~\eqref{eq:anisotropy}, where the $xy$ and $x^2-y^2$ orbitals have been swapped for the $x'y'z$ coordinate frame.
}
\begin{ruledtabular}
\begin{tabular}{c@{\hspace{2em}}rr@{\hspace{2em}}rr}
                    & \multicolumn{2}{c}{$xyz$} & \multicolumn{2}{c}{$x'y'z$}   \\\hline
	$\eps_{xy}$       &  $-0.892$ &               &  0.009     &                  \\
	$\eps_{yz}$       &  $-0.863$ & $A_x=81.6$\,K & $-0.802$   & $A_{x'}=87.7$\,K \\
	$\eps_{xz}$       &  $-0.834$ & $A_y=84.4$\,K & $-0.895$   & $A_{y'}=78.7$\,K \\
	$\eps_{3z^2-r^2}$ &  0.006    & $A_z=78.8$\,K &  0.006     & $A_z=78.8$\,K    \\
	$\eps_{x^2-y^2}$  &  0.009    &               & $-0.892$   &                  \\
\end{tabular}
\end{ruledtabular}
\end{table}
\subsubsection{Single-ion anisotropy}
The single-ion anisotropy term $AS_{i\varkappa}^2$, which is typical for Ni$^{2+}$, originates from weak distortions of the NiO$_6$ octahedron and depends on mutual positions of individual crystal-field levels~\cite{maekawa}. The contributions of the spin-orbit coupling $\lambda$ giving rise to the single-ion anisotropy can be obtained from second-order perturbation theory~\cite{maekawa}:
\begin{align}
  A_x= &\left(\frac{\lambda}{2}\right)^2\left(\frac{1}{\eps_{x^2-y^2}-\eps_{yz}}+\frac{3}{\eps_{3z^2-r^2}-\eps_{yz}}\right) \notag\\
	A_y= &\left(\frac{\lambda}{2}\right)^2\left(\frac{1}{\eps_{x^2-y^2}-\eps_{xz}}+\frac{3}{\eps_{3z^2-r^2}-\eps_{xz}}\right)\label{eq:anisotropy} \\
	A_z= &\left(\frac{\lambda}{2}\right)^2\left(\frac{4}{\eps_{x^2-y^2}-\eps_{xy}}\right), \notag
\end{align}
where $\lambda=0.078$\,eV is the spin-orbit coupling constant for Ni$^{2+}$~\cite{figgis1966}. The terms $A_{\gamma}$ enter the spin Hamiltonian as $\hat H=-\sum_{\gamma=x,y,z}A_{\gamma}S_{\gamma}^2$, such that the largest $A_{\gamma}$ determines the preferred spin direction. 

First, we use the standard $xyz$ coordinate frame, with the $z$ axis directed along the longer Ni--O bond (2.13\,\r A) and the $x$ and $y$ axes pointing along the shorter Ni--O bonds ($2.04-2.07$\,\r A). The resulting orbital energies listed in Table~\ref{tab:orbitals} do not follow crystal-field splitting anticipated for the octahedron elongated along $z$. For example, one expects $\eps_{xy}>\eps_{yz}\simeq\eps_{xz}$ because of the shorter Ni--O distances in the $xy$ plane, but we find $\eps_{xy}<\eps_{yz}<\eps_{xz}$ instead. 

A more natural model can be derived for the modified $x'y'z$ coordinate frame, where we turn the $x$ and $y$ axes by 45$^{\circ}$, so that they bisect Ni--O bonds in the $xy$ plane (Fig.~\ref{fig:structure}, left). This way, the $xy$ and $x^2-y^2$ orbitals are swapped, and their energies should be swapped in Eqs.~\eqref{eq:anisotropy} too. In the modified frame, we obtain $\eps_{y'z}>\eps_{x'z}\simeq\eps_{x'^2-y'^2}$ reflecting a peculiar scissor-like distortion of the NiO$_6$ octahedron~\cite{[{This effect closely resembles the scissor-like octahedral distortion in KTi(SO$_4)_2$, where an unusual orbital ground state of Ti$^{3+}$ is stabilized, see: }][{}]nilsen2015}. The $y'$ axis bisects the smaller O1--Ni--O2' angle of $86.45^{\circ}$ compared to the O1--Ni--O2 angle of $93.55^{\circ}$ bisected by the $x'$ axis. Therefore, $d_{x'z}$ is lower in energy than $d_{y'z}$. This relation has an immediate effect on the single-ion anisotropy. Using Eq.~\eqref{eq:anisotropy}, we find the effective anisotropy $A=A_{x'}-A_{y'}\simeq 9$\,K, and the direction of the easy axis $\varkappa=x'=[0.852,-0.023,-0.503]$ that bisects the two shorter \mbox{Ni--O} bonds of the elongated NiO$_6$ octahedron. Remarkably, this easy axis is consistent with the experimental spin direction $[0.89(9),-0.10(5),-0.49(6)]$ in the magnetically ordered state. Moreover, our microscopic estimate $A\simeq 9$\,K is compatible with the spin-flop field $H_{\rm SF}\simeq 9$\,T (Fig.~\ref{mh2}).

\section{Discussion}
\label{sec:ordered}
Li$_{2}$NiW$_{2}$O$_{8}$ reveals two successive magnetic transitions with two distinct magnetically ordered states. Taking into account the strong easy-axis anisotropy of Ni$^{2+}$, one would expect that these two states resemble magnetically ordered phases of CsNiCl$_3$~\cite{clark1972,kadowaki1987} and CsMnI$_3$~\cite{ajiro1990,harrison1991}, where a non-collinear (120$^{\circ}$-type) magnetic structure below $T_{\rm N2}$ is followed by a collinear state above $T_{\rm N2}$. Both states have the same commensurate propagation vector. 
Our NMR and neutron scattering experiments rule out this analogy for Li$_2$NiW$_2$O$_8$. We find commensurate magnetic ordering below $T_{\rm N2}$ and an incommensurate SDW type ordering between $T_{\rm N1}$ and $T_{\rm N2}$. 

Microscopically, Li$_2$NiW$_2$O$_8$ and its Ni$^{2+}$-based sibling CsNiCl$_3$ can be viewed as quasi-1D, with spin chains stacked on the triangular lattice. In CsNiCl$_3$, the leading intrachain coupling along $c$ manifests itself in characteristic excitations at higher energies~\cite{buyers1986,morra1988,zaliznyak1994}, whereas long-range magnetic order is essentially determined by the frustrated couplings in the $ab$ plane. The reduced magnetic moment in the ordered state (1.0\,$\mu_B$~\cite{yelon1973}) is taken as a hallmark of quantum fluctuations triggered by the 1D nature of the spin lattice~\cite{montano1972}. In Li$_2$NiW$_2$O$_8$, the ordered moment of 1.8\,$\mu_B$ approaches the full moment of 2\,$\mu_B$ for a spin-1 ion, thus indicating proximity to the 3D regime and minor role of quantum fluctuations. Indeed, the ratio of the intrachain to interchain couplings ($J_{01\bar 1}/\bar J_{ab}$) is as low as 3.8 in Li$_2$NiW$_2$O$_8$ compared to nearly 60 in CsNiCl$_3$~\cite{morra1988}.

CsNiCl$_3$ adopts a 120$^{\circ}$-like magnetic order below $T_{\rm N2}$ followed by a collinear order between $T_{\rm N2}$ and $T_{\rm N1}$ with the same, commensurate propagation vector. In contrast, magnetic order in Li$_2$NiW$_2$O$_8$ is collinear already below $T_{\rm N2}$. Antiparallel spin arrangement along $[100]$ and $[110]$, as opposed to the parallel spin arrangement along $[010]$, can be ascribed to the hierarchy of competing exchange couplings $J_{100}>J_{110}>J_{010}$, but collinear state itself is in fact not expected for the isotropic (Heisenberg) spin Hamiltonian on the triangular lattice, at least for the parameter regime derived in our work. 

Using exchange couplings from Table~\ref{tab:exchange}, we performed classical energy minimization and arrived at a spiral ground state with $\kv=(0.382,0.266,\frac12+k_y)$, where the incommensurability along the $c$ direction is due to the diagonal nature of the interplane coupling $J_{01\bar 1}$. This discrepancy between the classical spiral state and the collinear state observed experimentally has been reported for Li$_2$CuW$_2$O$_8$ as well~\cite{ranjith2015a}. Quantum fluctuations inherent to the spin-$\frac12$ Cu$^{2+}$ compound are a natural mechanism for stabilizing the collinear state, but in Li$_2$NiW$_2$O$_8$ the large ordered moment renders this possibility unlikely. On the other hand, the single-ion anisotropy $A\simeq 9$\,K favors the collinear state and overcomes the effect of frustration. It is worth noting that this mechanism is not operative in the Cu compound, where single-ion anisotropy vanishes for a spin-$\frac12$ ion.

The single-ion anisotropy is comparable to the leading exchange coupling, $A/J_{01\bar 1}\simeq 0.91$ in Li$_2$NiW$_2$O$_8$, while in CsNiCl$_3$ this ratio is 0.04 only~\cite{morra1988}. The pronounced anisotropy combined with the frustration on the triangular lattice may also be responsible for the formation of the incommensurate order between $T_{\rm N2}$ and $T_{\rm N1}$. The competition between the commensurate and incommensurate states is reminiscent of Ca$_3$Co$_2$O$_6$ that shows a peculiar sequence of magnetization plateaus in the applied magnetic field~\cite{kageyama1997,maignan2000,hardy2004}. Ca$_3$Co$_2$O$_6$ has been considered as a prototype material for spin chains stacked on the triangular lattice, although the actual geometry of its interchain couplings is more complex~\cite{chapon2009}. A striking feature of Ca$_3$Co$_2$O$_6$ is its incommensurate spin-density-wave state~\cite{agrestini2008,agrestini2008b} that is metastable and gradually transforms into commensurate magnetic order with time~\cite{agrestini2011}. Li$_2$NiW$_2$O$_8$ also shows an incommensurate-commensurate transformation, although both phases are thermodynamically stable and can be reversibly transformed upon heating or cooling. Importantly, though, Ca$_3$Co$_2$O$_6$ features ferromagnetic spin chains with a very strong single-ion anisotropy ($A/J\gg 1$)~\cite{jain2013,allodi2014,paddison2014}, while in Li$_2$NiW$_2$O$_8$ the spin chains are AFM, and $A/J\simeq 1$. 

\section{Summary and conclusions}
Li$_2$NiW$_2$O$_8$ reveals two successive magnetic transitions at $T_{\rm N1}\simeq 18$\,K and $T_{\rm N2}\simeq 12.5$\,K indicating two distinct magnetically ordered phases. Using nuclear magnetic resonance and neutron scattering, we have shown that the low-temperature phase is commensurate, with the propagation vector $\mathbf k=(\frac12,0,\frac12)$, while the intermediate-temperature phase is incommensurate, and its NMR signal resembles that of a spin-density wave. Collinear nature of the low-temperature phase is driven by the sizable single-ion anisotropy $A\simeq 9$\,K that is comparable to the leading exchange coupling $J_{01\bar 1}\simeq 9.9$\,K. 

The magnetic model of Li$_2$NiW$_2$O$_8$ can be represented by spin-1 chains stacked on the triangular lattice in the $ab$ plane. Despite the quasi-1D nature of the spin lattice with $J_{01\bar 1}/\bar J_{ab}\simeq 3.8$, the magnetic response of Li$_2$NiW$_2$O$_8$ is largely three-dimensional. We do not find broad maximum in the magnetic susceptibility above $T_{\rm N1}$, and the ordered magnetic moment $\mu=1.8(1)$\,$\mu_B$ at 1.5\,K is close to 2\,$\mu_B$ expected for a spin-1 ion, thus indicating minor role of quantum fluctuations in this compound.

1/$T_1$ below $T_{\rm N2}$ follows a $T^5$ behavior, indicating that the relaxation is mainly governed by three-magnon processes. Analysis of 1/$T_1$ in the critical regime just above $T_{\rm N1}$ suggests that the 3D ordering at $T_{\rm N1}$ is driven by the easy-axis anisotropy.

\acknowledgments
KMR and RN were funded by MPG-DST (Max Planck Gesellschaft, Germany and Department of Science and Technology, India) fellowship. AT was supported by the Federal Ministry of Education and Research through Sofja Kovalevskaya Award of Alexander von Humboldt Foundation. MS was funded by FP7 under Grant Agreement No. 290605 and by TRR80 of DFG. DK was supported by the DFG under FOR1346. We are grateful to PSI for providing the beamtime at DMC and acknowledge the support of the HLD at HZDR, member of EMFL. The usage of the HTE package~\cite{htse,*htsecode} is kindly acknowledged.

%

\end{document}